\documentclass[english,twocolumn,superscriptaddress]{revtex4-1}
\usepackage[T1]{fontenc}
\usepackage[latin9]{inputenc}
\usepackage{amsmath}
\usepackage{graphicx}
\usepackage{babel}

\begin{document}

\title{Thermodynamic efficiency of interactions in self-organizing systems}

\author{Ramil Nigmatullin}
\affiliation{Department of Physics and Astronomy, Macquarie University, NSW 2109, Australia.}
\affiliation{Centre for Complex Systems, Faculty of Engineering, The University of Sydney, Sydney, NSW 2006, Australia.}

\author{Mikhail Prokopenko}
\affiliation{Centre for Complex Systems, Faculty of Engineering, The University of Sydney, Sydney, NSW 2006, Australia.}
\date{September 2020}

\begin{abstract}
The emergence of global order in complex systems with locally interacting components is most striking at criticality, where small changes in control parameters result in a sudden global re-organization. We introduce a measure of thermodynamic efficiency of interactions in self-organizing systems, which quantifies the change in the system's order per unit work carried out on (or extracted from) the system. We analytically derive the thermodynamic efficiency of interactions for the case of quasi-static variations of control parameters in the exactly solvable Curie-Weiss (fully connected) Ising model, and demonstrate that this quantity diverges at the critical point of a second order phase transition. This divergence is shown for quasi-static perturbations in both control parameters, the external field and the coupling strength. Our analysis formalizes an intuitive understanding of thermodynamic efficiency across diverse self-organizing dynamics in physical, biological and social domains.
\end{abstract}

\maketitle

Typically, self-organization is defined as a spontaneous formation of  spatial, temporal, spatiotemporal structures or functions in a system comprising multiple interacting components. Importantly, a self-organizing process is assumed to be developing  in the absence of specific external controls, as pointed out by Haken \cite{haken}: 
\begin{quote}
\emph{a system is self-organizing if it acquires a spatial, temporal or functional structure without specific interference
from the outside. By `specific' we mean that the structure or functioning is not impressed on the system, but that the system is acted upon from the outside in a non-specific fashion. For instance, the fluid which forms hexagons is heated from below in an entirely uniform fashion, and it acquires its specific structure by self-organization.}
\end{quote} 
To explain structures that spontaneously
self-organize when energy or matter flows into a system typically describable by many variables, Haken employed the notion of {\em order parameters} (degrees of freedom) and control parameters \cite{synerg,haken}: slowly varying a relevant control parameter, such as temperature of a ferromagnetic material, may induce an abrupt
change, a {\em phase transition}, in an observable order parameter, such as the net magnetization. 
The emergence of global order in complex systems is most striking
at criticality, when the characteristic length and dynamical time
scales of the system diverge. A phase transition is usually accompanied by global symmetry breaking.
Crucially, in the more organized (coherent) phase of the system dynamics, the global behavior of the system can be described by only a few order parameters, that is, the system becomes low-dimensional as some dominant variables ``enslave'' others. A canonical physical example of such coherent dynamics, when the whole system acts in synchrony, is laser: a beam of coherent light created out of the chaotic movement of particles \cite{haken}. 

In physical systems, the local interactions are usually determined by physical laws, e.g., interactions among fluid molecules or crystal ions, while the interactions within a biological organism may evolve over generations under environmental selection pressures, bring survival benefits. The role of locally interacting particles contributing to self-organizing pattern formation in biological systems has been captured in a definition offered by Camazine {\em et al.}~\cite{camaz}: 
\begin{quote} \emph{Self-organization is a process in which pattern at the global level of a system emerges solely from numerous interactions among the lower-level components of the system. Moreover, the rules specifying interactions among the system's components are executed using only local information, without reference to the global pattern.}
\end{quote} 
These definitions concur with many other approaches to formalize self-organization, highlighting three important aspects \cite{bona,polani2003,Prokopenko2008}: (i) a system dynamically advances to a more organized state, while exchanging energy, matter and/or information with the environment, but without a specific external ordering influence; (ii) the interacting system components have only local information, and so exchange only local information, but exhibit long-range correlations; (iii) the increase in organization can be observed as a more coherent global behavior.

In general, as the state of a complex system evolves, its configurational entropy changes. The reduction (or increase) in the configurational entropy occurs at the
expense of work extracted or carried out on the system, and the heat exported to the environment. Thus, a thermodynamic analysis of the interactions in self-organizing systems aims to quantify the work, heat and energy exchange between the system and the environment. 
One can reasonably expect that self-organization is most thermodynamically efficient in the vicinity of the critical points, i.e., at criticality one may expect that a smaller amount of work extracted/done on a system can result in a larger change of the configurational entropy. 
Indeed, it has been conjectured before that a system in a self-organized low-dimensional phase with fewer available configurations (i.e., describable by just a few order parameters and exhibiting macroscopic stability) may be more efficient than the system in a high-dimensional disorganized phase with more configurations.

To formalize this conjecture, Kauffman proposed a succinct principle behind the higher efficiency of self-organized systems --- the generation of constraints during the release of energy --- the constrained release channels energy to perform some useful work, which can propagate and be used again to create more constraints, releasing further energy and so on \cite{Kau00}. Following a similar characterization, Carteret {\em et al.}~\cite{Carteret} have shown that available power efficiency is maximized at \emph{critical} Boolean networks. The question of thermodynamic efficiency has also been proposed and studied in the context of the cellular information-processing, from the perspective of how close life has evolved to approach maximally efficient computation ~\cite{barato2014efficiency,Kempes2017}. Furthermore, a recent thermodynamic analysis of a model of active matter  demonstrated that the efficiency of the collective motion diverges at the transition between disordered and coherent collective motion \cite{Crosato2018Thermodynamics}. However, the precise nature of the divergence of the efficiency of collective motion, and its relation to the critical exponents describing the system behavior in the vicinity of the phase transitions remained unclear, due to the lack of  analytical expressions for the corresponding configurational probability distributions.

In this study, we offer a generic measure of thermodynamic efficiency of interactions within self-organizing systems, aiming to clearly differentiate between phases of system dynamics, and identify the regimes when the efficiency is maximal. This measure is expressed by contrasting (i) the change of organization attained within the system (i.e., change in the created order or predictability) with (ii) the thermodynamic work involved in driving such a change. We demonstrate that the maximal efficiency is indeed achieved \emph{at the critical regime}, i.e., during the phase transition, rather than at the macroscopically stable low-dimensional phase \emph{per se}. The reasons for the maximal efficiency exhibited by systems during self-organization, i.e., at a critical regime, are articulated precisely in terms of the increased order (or the reduction of Shannon entropy) related to the amount of the  work carried out during the transition. This measure is defined for specific configurational changes (perturbations), rather than states or regimes --- in line with the point made by Carteret {\em et al.}~\cite{Carteret} that the maximization of power efficiency occurs at a finite displacement from equilibrium.

In developing our approach we build on information-theoretic and statistical-mechanical methods, interpreting the process of self-organization as a thermodynamic phenomenon, while considering the interactions within the system as distributed information processing or distributed computation~\cite{Carteret,barato2014efficiency,Prokopenko-Einav,Spinney2016-arrow,Kempes2017,Crosato2018Thermodynamics}. 
Our aim is to develop a common understanding of thermodynamic efficiency across multiple examples of self-organizing dynamics in physical, biological and social domains. These phenomena include transitions from disordered to coherent collective motion~\cite{Gregoire2004,Buhl2006,Vicsek2006,Szabo2007,mora2011biological,bialek2012statistical,Crosato2018Thermodynamics},  phase transitions in spin systems and active matter \cite{Montanari,FTE,Crosato2019}, chaos-to-order transitions in genetic regulatory networks modeled as random Boolean networks~\cite{Carteret,Wang-Fisher},  synchronization in networks of coupled oscillators near ``the edge of chaos''~\cite{Kuramoto1987StatisticalMO,Miritello,Kalloniatis2018},   transitions across epidemic thresholds during contagions and cascading failures~\cite{Newman1999,Sander2002,Cupac2013,WangLiu2015,Harding2018}, critical dynamics of urban evolution~\cite{WilsonDearden,Crosato2018Urban,Slavko2019}, among many others. Self-organizing criticality (SOC)~\cite{SOC} is a related but distinct phenomenon, as we are not attempting to reveal the mechanisms of self-organization towards critical regimes, focusing instead on defining and determining the thermodynamic efficiency of interactions in a representative self-organizing system.

In this work, we select an abstract statistical-mechanical model (Curie-Weiss model of interacting spins in a fully connected graph) --- one of the simplest model exhibiting a second-order phase transition --- from the widely applicable mean-field universality class. We analytically evaluate dynamics of this model in the vicinity of a phase transition, and prove that the thermodynamic efficiency  has a power law divergence at the critical point, and compute its critical exponent.

\section{Framework} 

Consider a statistical mechanical system in thermodynamic
equilibrium, where $\textbf{X}=\{X_{1},...,X_{n}\}$ are intensive thermodynamic quantities which act as control
parameters that can be changed externally (e.g., magnetic field).
A perturbation in the control parameter, $\textbf{X}\rightarrow\textbf{X}+\delta\textbf{X}$,
will result in a change in thermodynamic potentials in the system
including its entropy and energy. We define the thermodynamic efficiency
of interactions as 
\begin{equation}
\eta(\textbf{X};\delta\textbf{X})=\frac{1}{k_{B}}\frac{\delta S}{\delta W},\label{eq:eta}
\end{equation}
where $\delta S$ and $\delta W$ are the change in entropy and the
work done/extracted on the system due to the perturbation $\delta\textbf{X}$.
Entropy $S$ is a configurational entropy, and thus $\eta(\textbf{X};\delta\textbf{X})$
quantifies the reduction (increase) of uncertainty in the state of the system
that we gain per unit of work done. A high value of $\eta$ signifies
that it is energetically easy to create order (reduce the configurational
uncertainty) in the system by changing a control parameter, whereas
a low value of $\eta$ indicates that a lot of work is needed to change the
order in the system. 

In practice, to evaluate $\eta(\textbf{X};\delta\textbf{X})$ we need
to specify the perturbation protocol. A change in control parameters
moves the system out of thermal equilibrium, and we need to compute
the amount of work done/extracted, $\delta W$, as the system relaxes back to
its equilibrium state. Thus, $\eta(\textbf{X};\delta\textbf{X})$
depends on how we perturb the system, and on the master equation that
describes the relaxation of the system back to its equilibrium state.
In what follows we will consider the case of a quasi-static perturbation
protocol, i.e., we  assume that the perturbation is sufficiently
slow that the system effectively adjusts instantaneously to its new
equilibrium state. 
Helmholtz free energy, $F(\theta,\textbf{X})$, is the most useful
thermodynamic potential for analyzing the quasi-static protocols at
constant temperature. Helmholtz free energy is related to the internal
energy $U$ and entropy $S$ via equation 
\[
U(\theta,\textbf{X})=\theta S(\theta,\textbf{X})+F(\theta,\textbf{X}),
\]
where $\theta\equiv k_{B}$T. To a first order in $\delta\textbf{X}$
the change in internal energy, entropy and free energy induced by
varying the control parameters are $\delta U=\delta\textbf{X}\cdot\left.\nabla U\right|_{\textbf{X}}$, $\delta S=\delta\textbf{X}\cdot\left.\nabla S\right|_{\textbf{X}}$
and $\delta F=\delta\textbf{X}\cdot\left.\nabla F\right|_{\textbf{X}}$. In a quasi-static process the change in free energy can be identified
with the work done on the system, $\delta F=\delta W$, and the entropy
change in the system balances the entropy exported to the environment,
$\delta S=-\delta S^{\textrm{exp}}$. Thus, for a quasi-static protocol,
the thermodynamic efficiency reduces to 
\begin{equation}
\eta(\textbf{X};\delta\textbf{X})=\frac{1}{k_{B}}\frac{\delta\textbf{X}\cdot\left.\nabla S\right|_{\textbf{X}}}{\delta\textbf{X}\cdot\left.\nabla F\right|_{\textbf{X}}}.\label{eq:eta-1}
\end{equation}

In the case when the variation of control parameter is one-dimensional
$\textbf{X}=X$, equation (\ref{eq:eta-1}) simplifies to 
\begin{align}
\eta(X, \delta X) & = \left. \frac{1}{k_{B}}\frac{\partial S}{\partial X} \middle/ \frac{\partial F}{\partial X}\label{eq:eta-2} \right. \\
 & =\frac{1}{k_{B}}\frac{\partial S}{\partial F}.\nonumber 
\end{align}

Equation (\ref{eq:eta-2}) applies for general quasistatic processes. When the system is close to a critical point of a phase transition, expression for $\eta$ can further be simplified using the following argument. Let $\psi$ be an extensive quantity conjugate to $X$ 
\begin{equation}
    \psi = - \frac{\partial F}{\partial X}.
\end{equation}

Entropy is related to free energy via $S=-\partial F/\partial T$, and thus the derivative of $S$ with respect to $X$ is 

\begin{equation*}
    \frac{\partial S}{\partial X} = - \frac{\partial^2 F}{\partial X \partial T} = - \frac{\partial^2 F}{\partial T \partial X} = \frac{\partial \psi}{\partial T}.
\end{equation*}

Thus, in terms of the extensive variable conjugate to the control parameter the efficiency given by equation (\ref{eq:eta-2}) can be expressed as 

\begin{equation}
    \eta(X,\delta X) = -\frac{1}{k_B} \frac{1}{\psi} \frac{\partial \psi}{\partial T} \label{eq:eta_ext}
\end{equation}

If $\psi$ is an order parameter of a phase transition then near the critical point we have $\psi = a |T-T_c|^\beta$, where $T_c$ is the critical temperature, $\beta$ is the critical exponent and $a$ is non-universal proportionality constant. Upon substitution of this expression for $\psi$ into (\ref{eq:eta_ext}), the constant $a$ cancels and we obtain 

\begin{equation}
    \eta(X,\delta X) = -\frac{1}{k_B} \frac{\beta}{|T-T_c|}. \label{eq:eta_critical}
\end{equation}

Equation (\ref{eq:eta_critical}) expresses the divergence of $\eta$ solely in terms of universal exponent $\beta$. This result explains why in many thermodynamic models the efficiency of self-organization is expected to peak near the critical point.

In many complex systems, there may not exist a readily available physical model expressed in terms of a Hamiltonian and the expression for the order parameter may not be evident. Nevertheless, if there is a record of samples of the states of the system then one may still use equation (\ref{eq:eta-2}) to compute the efficiency of self-organization. The reason is that all of the thermodynamic quantities in (\ref{eq:eta-2}), expressed in
terms of Gibbs probability distribution, have a clear information-theoretic
interpretation. Entropy $S$ is directly proportional to the Shannon
entropy $H$, $S=-k_{B}\sum_{x}p(x)\log p(x)=k_{B}H$. The free energy
$F$ is related to the Fisher information $\mathcal{I}$ via equation $\mathcal{I}=\partial F^{2}/\partial^{2}X$,
with the Fisher information quantifying the sensitivity of the probability
distribution to the change in the control parameter, $\mathcal{I} \equiv \sum_{x}(\partial\log p(x)/\partial X)^{2}p(x)$.

There are several interpretations of the Fisher information relevant to critical dynamics and scale dependence: $\mathcal{I}$ is equivalent to the thermodynamic metric tensor, a curvature of which diverges at phase transitions; $\mathcal{I}$ measures the size of the fluctuations in the collective variables around equilibrium; also, $\mathcal{I}$  is proportional to the derivatives of the corresponding order parameters with respect to the collective variables \cite{brody1995geometrical,brody2003information,janke2004information,crooks2007measuring,Wang-Fisher,prokopenko2011relating,Machta2013}. Substituting $\partial F/\partial X = \int \mathcal{I} dX$ into equation (\ref{eq:eta-2}) gives
\begin{equation}
\eta(X)=\frac{\partial H/\partial X}{\int\mathcal{I}dX}.\label{eq:eta-3}
\end{equation}
Equation (\ref{eq:eta-3}) expresses the thermodynamic efficiency of interaction during configurational perturbations in terms of information-theoretic quantities of entropy and Fisher information.

Equation (\ref{eq:eta-3})
was derived and used in \cite{Crosato2018Thermodynamics} in the thermodynamic analysis
of collective motion (e.g., swarming) exhibiting a kinetic phase transition. Crosato et al.~\cite{Crosato2018Thermodynamics} computed the efficiency $\eta$ from the distribution $p(x)$,
%$p(x;X)$ 
 estimated via sampling produced by numerical simulations of the model, consequently yielding estimates of $H$
and $\mathcal{I}$. It was then demonstrated that $\eta$ diverges
at the critical point where the swarm transitions from disordered 
to coherent motion. 

The notion of thermodynamic efficiency $\eta$ was also applied to the analysis of urban transformations~\cite{Crosato2018Urban}, driven by quasi-static changes in the social disposition: a control parameter 
characterising the attractiveness of different areas.  The thermodynamic efficiency of urban transformations was defined as the reduction of configurational entropy resulting from the expenditure of work. In the socioeconomic context of urban dynamics, it expressed the ratio of the gained predictability of income flows to the amount of work required to change the social disposition. Importantly, the efficiency was shown to peak at a critical transition separating dispersed and polycentric phases of urban dynamics~\cite{Crosato2018Urban}. 

Similarly, Harding et al.~\cite{Harding2018} considered thermodynamic efficiency of quasi-static epidemic processes, defined for a value of some control parameter (e.g., the infection transmission rate), as the ratio of the reduction in uncertainty to the expenditure of work needed to change the parameter. On the one hand, this could be the efficiency of an intervention process consuming work in order to reduce the transmission rate. On the other hand, the efficiency can be defined in terms of the pathogen emergence --- a process which increases the transmission rate, and in doing so extracts the work. Irrespective of the interpretation, the efficiency was shown to peak at the epidemic threshold~\cite{Harding2018}.

Our contribution builds on this research, showing that according to equation (\ref{eq:eta_critical}), the divergence of the efficiency of self-organization is generally expected to occur at a second-order phase transition. In the following section, we illustrate this result by explicitly computing $\eta$ in a simple model exhibiting paramagnetic to ferromagnetic phase transition, showing that the efficiency of self-organization peaks at the critical point when the control parameter is either the coupling strength between the spins or the external magnetic field.

\section{Example: Curie-Weiss Model}

The energy of a wide variety statistical mechanical systems, including spin glasses, can be written in the following form

\begin{equation}
E(\underline{\sigma},\{X_i\})= E_0 +\sum_i X_i E_i(\underline{\sigma}),\label{eq:statmech}
\end{equation}
where $\{X_i\}=\{X_1,X_2,...,X_K\}$ are the control parameters of the system, $\underline{\sigma}$ denotes the microscopic state of the system and $\theta\equiv k_B T$. We are working with a canonical ensemble, where the system is in contact with a heat bath in thermal equilibrium and the average energy is fixed. In this case, the probability of finding the system in configuration $\underline{\sigma}$
is given by the Gibbs measure
\begin{equation}
p(\underline{\sigma};\{X_i\})=\frac{e^{-E(\underline{\sigma},\{X_i\})/\theta}}{Z(\theta,\{X_i\})},\label{eq:prob}
\end{equation}
where $Z=\sum_{\underline{\sigma}} e^{-E(\underline{\sigma})}/\theta$ is the partition function. The free energy of the system is given by $F=\ln Z$. The free energy 
The free energy can be used to compute any thermodynamic quantity, in particular, the expectation of $\langle E_1 \rangle, ...,\langle E_K \rangle$ are given by

\begin{equation}
    \langle E_j \rangle = -\frac{\partial F}{\partial X_j}.
\end{equation}

For an interacting statistical mechanical system in thermal equilibrium there is a one to one map between the the set of control parameters $\{T,X_1,...,X_K\}$ and $\{S,\langle E_1\rangle,...\langle E_K\rangle \}$ \cite{Wei2016}, and thus we will refer to $\psi_i \equiv \langle E_i \rangle$ as a order parameter conjugate to the control parameter $X_i$. Phase transitions are often accompanied by divergences in one or more order parameters $\psi_i$ or their derivatives. 

In the rest of the paper, we will focus on computing $\eta$ for a system governed by a specific energy function of the form (\ref{eq:statmech}) --- the Curie-Weiss (CW) model. The CW model is a model of ferromagnetism, where each spin interacts with all other spins via pairwise interactions and for this reason it is also known as the fully connected Ising model. This model exhibits a second-order phase transition at
a finite critical temperature $T_{c}$. In the vicinity of the critical
point, the analytic expression to all of the thermodynamic
quantities are known, which enables the derivation of the analytic
expression for $\eta$. The phase transition from ferromagnetic to paramagnetic states in the Curie-Weiss model belongs to the mean field universality class.  

Let $N$ spins $\sigma_{i}\in\{\pm1\}$ be assigned
to sites $i\in\{1...N\}$. A configuration of the system is given
by $\underline{\sigma}=(\sigma_{1},...,\sigma_{N})$. The energy function
for the system containing pairwise interactions between spins and
in the presence of an external magnetic field $B$ is given by 
\begin{equation}
E(\underline{\sigma})=-\frac{J}{N}\sum_{(ij)}\sigma_{i}\sigma_{j}-\mu B\sum_{i=1}^{N}\sigma_{i},\label{eq:E}
\end{equation}
where the sum over $(ij$) runs over all of the $N(N-1)/2$. The $1/N$ scaling in front of the spin-spin
interaction term is to yield an extensive free energy. 
In this model, the control parameters are $\{J, B\}$, which respectively denote exchange interaction strength and externally applied magnetic field.
The probability of finding the system in configuration $\underline{\sigma}$
is given by the Gibbs measure
\begin{equation}
p(\underline{\sigma};T,B,J)=\frac{e^{-E(\underline{\sigma})/\theta}}{Z_{N}(\theta,B,J)},\label{eq:prob}
\end{equation}
where $\theta\equiv k_{B}T$ and $Z_{N}$ is a partition function
for the $N$-spin system. The free energy of the $N$ spin system
is given by $F_{N}(\theta,B)=\ln Z_{N}(\theta,B)$. The thermodynamic
limit is obtained by taking $N\rightarrow\infty$. In the thermodynamic
limit the free energy density $f(\theta,B)=\lim_{N\rightarrow\infty}F_{N}(\theta,B)/N$
can have the following analytic expression \cite{CWmodel_paper}:
\begin{equation}
f(\theta,B)=-\theta\ln2-\theta\ln(\Phi(\theta,B))\label{eq:f_dens}
\end{equation}
with 
\begin{equation}
\Phi(\theta,B)=e^{-Jy^{2}/(2\theta)}\cosh\left(\frac{Jy+B}{\theta}\right)\label{eq:Phi}
\end{equation}
Here $y$ is defined as solution to the equation 
\begin{equation}
y=\tanh\left(\frac{Jy+\mu B}{\theta}\right).\label{eq:y}
\end{equation}

The average magnetization per spin is the order parameter conjugate to magnetic field and is given by $m=-(\partial f/\partial B)_{\theta}=\mu y$, and thus the
equation of state is $m=\mu\tanh[(Jm+B\mu)/\theta\mu]$. The phase
diagram can be constructed by analyzing the equation of state. The
critical point of a second-order phase transition occurs at $B=0$
and $\theta_{c}=J$. When $B=0$ and $\theta>J$ there is only one stable
solution of the equation of state, which is $m=0$. When $B=0$ and
$\theta<J$ , there are three solutions: one unstable solution $m=0$
and two stable solution $m=\pm m^{*}$ where $m^{*}$ is found by numerically
solving the equation $m=\mu\tan(Jm/\theta)$. Thus at $B=0$ and at
the critical temperature $\theta_{c}=J$, the system transitions from a paramagnetic
disordered state where $m=0$ to a ferromagnetic ordered state where
$m=\pm m^{*}$. This transition is of second order, since the second
derivatives of $f$ with respect to both $B$ and $\theta$ are discontinuous at $\theta_{c}$. 

Having reviewed the phase change behavior of the CW model,
we will now evaluate the thermodynamic efficiency 
$\eta$ associated with varying the magnetic field $B$ along a quasi-static
protocol. The entropy density is related to the free energy density
via equation
\begin{equation}
s=-\frac{\partial f(y(\theta,B),\theta,B)}{\partial\theta}.\label{eq:entropy}
\end{equation}

Using the equations (\ref{eq:f_dens})-(\ref{eq:entropy}) and (\ref{eq:eta-1}), one can compute efficiency of self-organization $\eta$ resulting from variation one or more control parameters $B$, $J$ or $\theta$.

\subsection{Varying external field, $B$}
\begin{figure}
\centering

\includegraphics[scale=0.8]{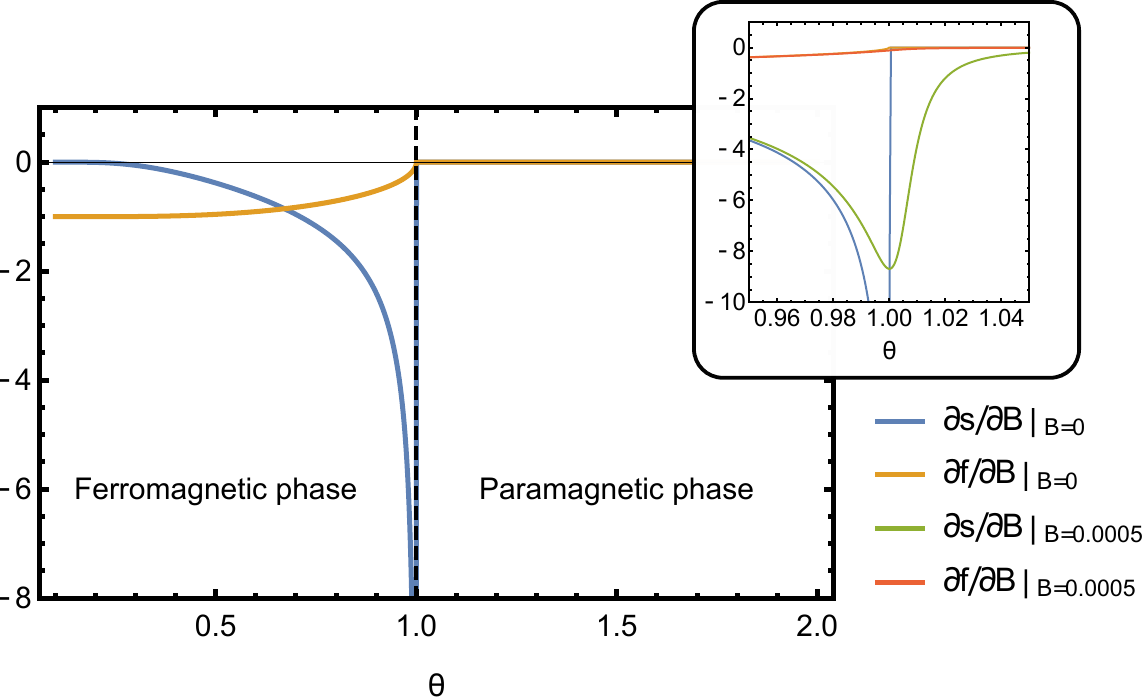}

\caption{Derivatives of entropy and free energy as a function of temperature
$\theta$ at zero magnetic field. The inset shows how the presence
of a small magnetic field smooths out the singularity in $\partial S/\partial B$
at the critical point $\theta_{c}=J=1.$ \label{fig:Derivatives}}
\end{figure}

\begin{figure}
\centering

\includegraphics[scale=0.74]{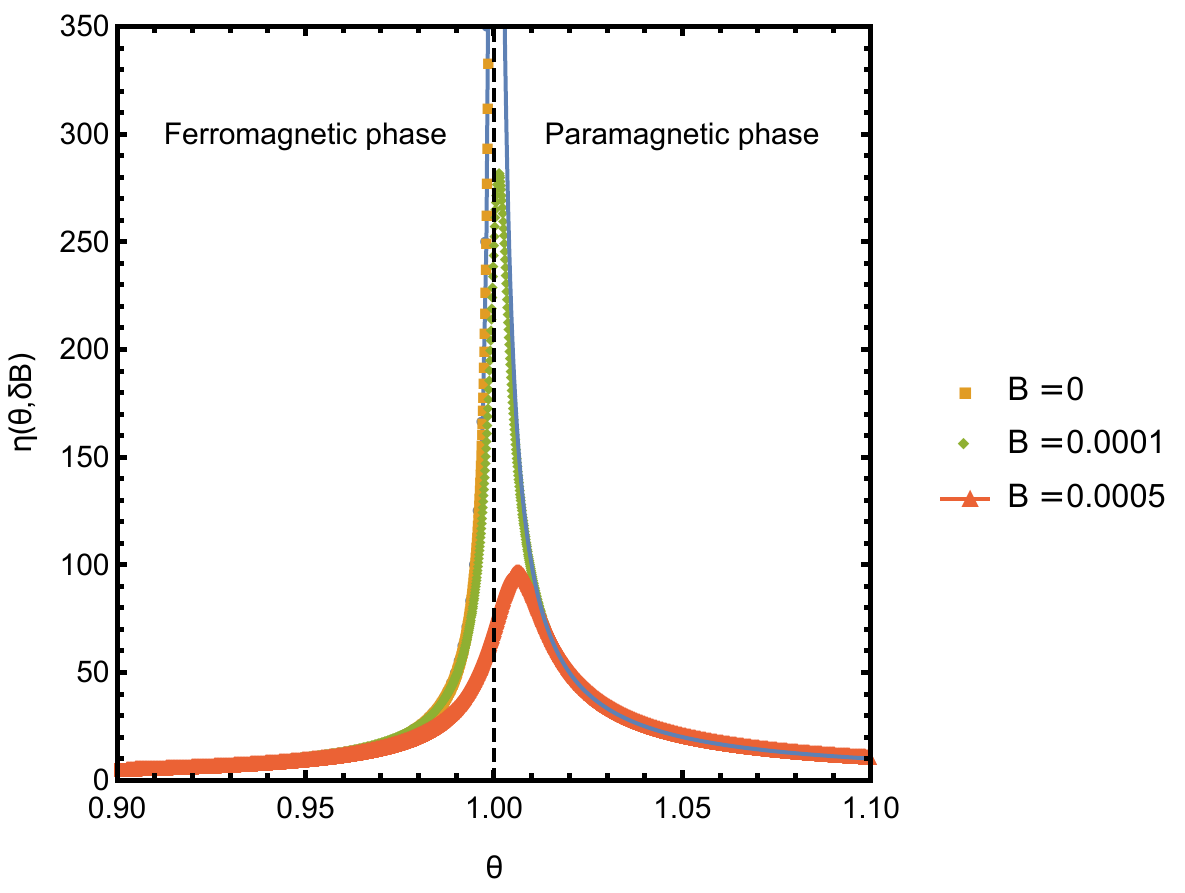}

\caption{Thermodynamic efficiency  $\eta(\theta,\delta B)$
as a function of $\theta$ at several small values of $B$. The critical point is at $\theta_c = 1.0$ or equivalently at $t\equiv (\theta-\theta_c)/\theta_c=0$.   For $t>0$,
$\eta$ is undefined at $B=0$. The solid lines $-1/2t^{-1}$ for
$t<0$ and $t^{-1}$ for $t>0$ are analytic expressions for $\eta$
in the vicinity of the critical point. \label{fig:eta}}
\end{figure}

Since equation (\ref{eq:y}) does not have a closed form solution
for $y(\theta,B)$, it has to be solved numerically. Thus, for a general choice of parameters of the Curie-Weiss model, the efficiency $\eta(\theta;\delta B)$ needs to be evaluated numerically. The plots of derivatives of free energy and entropy densities computed numerically by solving equation (\ref{eq:y}) are shown in Figure \ref{fig:Derivatives}.
The thermodynamic efficiency is the ratio of
these two derivatives, $\eta=\frac{1}{k_{B}}\frac{\partial s}{\partial B}/\frac{\partial f}{\partial B}$, which is plotted in Figure \ref{fig:eta}. As expected, the efficiency $\eta$ peaks near the critical point of the phase transition. In the rest of this section, we focus on behaviour of $\eta$ in the vicinity of the critical point, where it is possible to obtain an analytic solution for all thermodynamic quantities and study their scaling behaviour.     

Near the critical point of the paramagnetic to ferromagnetic phase transition, the average magnetization $y/\mu$ is small and thus equation of state (\ref{eq:y}) can be approximated by a low order Taylor expansion in $y$. Keeping up to $O(y^{3})$ the equation of state is 
\begin{equation}
K^{3}y^{3}-3y(K-1)-3h=0,\label{eq:eq_state2}
\end{equation}
where $K\equiv J/\theta=\theta_{c}/\theta$, $h=\mu B/\theta$. In
the case of zero magnetic field, $h=0$, the solution of (\ref{eq:eq_state2})
is 
\begin{align}
y & =0,\qquad\textrm{for }t\geq0,\label{eq:magnetization-1}\\
 & =\pm\sqrt{\frac{3(K-1)}{K^{3}}}\sim\sqrt{3}\left(-t\right)^{1/2}\qquad\textrm{for }t<0\nonumber 
\end{align}
where $t$ is the reduced temperature $t\equiv(\theta-\theta_{c})/\theta_{c}$
and $h\equiv\mu B/\theta$. Equation (\ref{eq:eq_state2}) produces
the well-known mean field scaling law for magnetization $m\sim(-t)^{\beta}$
for $t<0$, with the critical exponent $\beta=1/2$. Use equation (\ref{eq:eta_critical}) we arrive at $\eta(\theta, \delta B) = -\frac{1}{2 k_B} \frac{1}{t}$ for $t < 0$.

In the paramagnetic case $t\geq0$, $y=0$, both $\partial f/\partial B$ and $\partial s/\partial B$ are zero and consequently the efficiency appears to be undefined as it is a ratio of these derivatives. Nevertheless, in the paramagnetic regime the derivatives of free energy and entropy can be made finite by either adding a small external magnetic field or by considering a finite size system. Here, we will consider the efficiency $\eta$ in the presence of constant magnetic field $B_0$, which can be made arbitrary small. In the presence of external field and when $t\gg 0$, the equation of state (\ref{eq:eq_state2}) simplifies to $y(1-K)+h=0$,
since the term $K^{3}y^{3}$ is negligible. Thus, in this regime, $y\sim h/(1-K)= \mu B/(\theta-\theta_c)$
and $f$ can now be evaluated using equations (\ref{eq:f_dens}) and
(\ref{eq:Phi}). From $f$ we compute $\partial_{B}f$, $\partial_{B}s$ and then Taylor expand to the leading order in $B$ to obtain
\begin{align}
\left.\frac{\partial f}{\partial B}\right|_{B=B_{0}} & =-\frac{B_{0}}{\theta_c t}\qquad\textrm{for }t\geq0 \label{eq:dfdb1}\\
\left.\frac{\partial s}{\partial B}\right|_{B=B_{0}} & =-\left.\frac{\partial f}{\partial B \partial \theta}\right|_{B=B_{0}}=-\frac{B_0}{\theta_c^2 t^2}\qquad\textrm{for }t\geq0. \label{eq:dsdb1}
\end{align}

Now, we can evaluate scaling behavior of the thermodynamic efficiency
 $\eta$ around the critical point:
\begin{align}
\eta(\theta,\delta B) & = \left. \frac{1}{k_{B}}\frac{\partial s}{\partial B} \middle/ \frac{\partial f}{\partial B} \nonumber \right.\\
= & \begin{cases}
-\frac{1}{k_{B}}\frac{1}{2}t^{-1} & \qquad\textrm{for }t<0\\
\frac{1}{k_{B} \theta_c}t^{-1} & \qquad\textrm{for }t>0.
\end{cases}\label{eq:eta_scaling}
\end{align}

A plot of $\eta$ in the vicinity of the critical point for several
small values of bias field $B_{0}$ is shown in figure \ref{fig:eta}.
The curves were obtained by numerically solving for $y$ and numerically
computing the derivative of $f$ and $s$. The $|t|^{-1}$ scaling
prediction agrees very well with the numerical results. The deviations
at finite $B_{0}$ and very close to the critical point are expected
as the scaling was obtained by neglecting $K^{3}y^{3}$ term in equation
(\ref{eq:eq_state2}), which is not small around $\theta=\theta_{c}$. 

\subsection{Varying coupling strength, $J$}

We now consider computing $\eta$, when $J$ is used as a control parameter. In this case, the relevant order parameter is $\psi_J \equiv \langle \sum_{ij} \sigma_i \sigma_j \rangle$, which quantifies the interaction energy between pairs of spins.
The spins will spontaneously align at a critical value of coupling
strength $J_{c}=\theta$, and the efficiency  $\eta(J,\delta J)$
is expected to peak near the critical point. 

Near the critical point
there is a closed form expression for $y$, and thus we can derive
the scaling relation between $\eta$ and the reduced coupling strength
$\mathcal{J}\equiv(J-J_{c})/J_{c}$. For the ferromagnetic case, $J>J_{c}$,
inserting equation (\ref{eq:magnetization-1}) into the expressions
for the free energy and entropy, taking derivatives with respect to
$\mathcal{J}$ and then Taylor-expanding in $\mathcal{J}$ to the lowest orders yields
\begin{align}
\frac{\partial f}{\partial \mathcal{J}} & =-\frac{3 \mathcal{J}}{2\theta}\qquad\textrm{for }\mathcal{J}>0\label{eq:f2}\\
\frac{\partial s}{\partial \mathcal{J}} & =-\frac{3}{2\theta}\qquad\textrm{for }\mathcal{J}>0\label{eq:s2}
\end{align}

The order parameter conjugate to $\mathcal{J}$ can be defined as $\phi\equiv -\partial f/\partial \mathcal{J}$, which according to equation (\ref{eq:f2}) is linearly proportional to $\mathcal{J}$, i.e. $\phi\sim\mathcal{J}^\beta$ with $\beta=1$. Using equation (\ref{eq:eta_critical}) with critical exponent $\beta=1$, we immediately arrive at $\eta(\theta, \delta \mathcal{J}) = \frac{1}{ k_B} \frac{1}{\mathcal{\mathcal{J}}}$ for $\mathcal{J} > 0$.

In the paramagnetic case, $\mathcal{J}<0,$ the magnetization is zero in the
absence of the external magnetic field and the efficiency of interactions is
undefined since $\partial f /\partial \mathcal{J} =0$. However, in the presence of small
bias magnetic field $B_{0}$, $\eta$ can be computed, since in that
case $y\sim h/(1-K)=\mu B_{0}/(\theta-J)$. Taylor-expanding $\partial f /\partial \mathcal{J}$
and $\partial s / \partial \mathcal{J}$ computed with this expression for $y$ to the
lowest orders in $\mathcal{J}$ gives 
\begin{align}
\left.\frac{\partial f}{\partial \mathcal{J}}\right|_{B=B_0} & =-\frac{B_0^{2}\mu^{2}}{\mathcal{J}^{3}}\qquad\textrm{for }\mathcal{J}<0\label{eq:f3}\\
\left.\frac{\partial s}{\partial \mathcal{J}}\right|_{B=B_0} & =\frac{2B_0^{2}\mu^{2}}{\mathcal{J}^{4}}\qquad\textrm{for }\mathcal{J}<0.\label{eq:s3}
\end{align}
Using equations (\ref{eq:f2})-(\ref{eq:s3}) we can compute the efficiency
of interactions in the vicinity of the critical point: 
\begin{align}
\eta(J,\delta J) & = \left. \frac{1}{k_{B}}\frac{\partial s}{\partial \mathcal{J}} \middle/ \frac{\partial f}{\partial \mathcal{J}}\nonumber \right. \\
= & \begin{cases}
\frac{1}{k_{B}}\mathcal{J}^{-1} & \qquad\textrm{for }\mathcal{J}>0 \\
-\frac{1}{k_{B}}2\mathcal{J}^{-1} & \qquad\textrm{for }\mathcal{J}<0.
\end{cases}\label{eq:eta_scaling-1}
\end{align}

\begin{figure}
\centering

\includegraphics[scale=0.74]{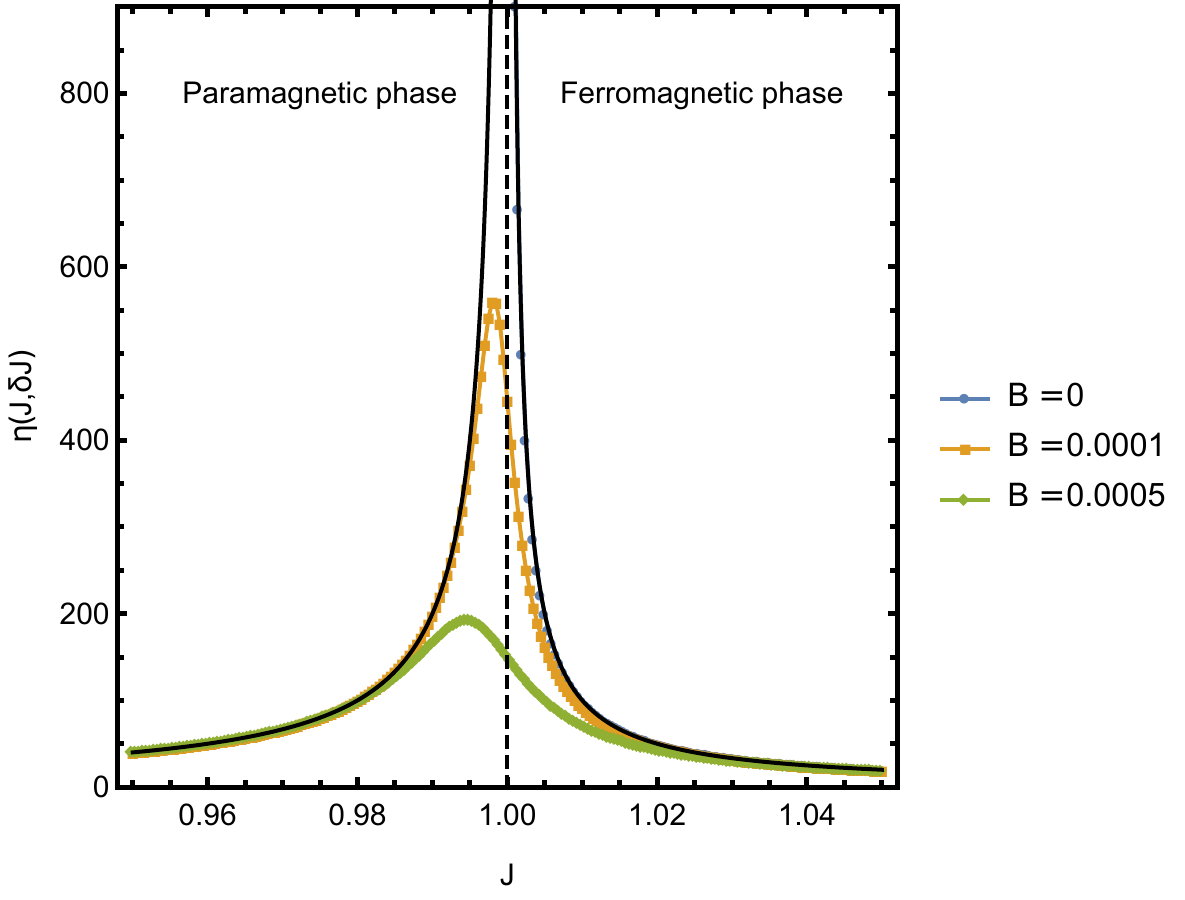}

\caption{Thermodynamic efficiency  $\eta(J,\delta J)$
as a function of $J$ at several small values of $B$ at $\theta=1.0$.
The critical point is at $J_{c}=1.0$ or equivalently at $\mathcal{J}\equiv(J-J_{c})/J_{c}=0$.
For $\mathcal{J}<0$, $\eta$ is undefined at $B=0$. The solid lines $-2\mathcal{J}^{-1}$
for $\mathcal{J}<0$ and $\mathcal{J}^{-1}$ for $\mathcal{J}>0$ are analytic expressions for $\eta$
in the vicinity of the criticality. \label{fig:eta-1}}
\end{figure}

Figure \ref{fig:eta-1} shows the plot of $\eta(J,\delta J)$ in the
vicinity of the critical point for several small values of bias field
$B_{0}$. The dotted curves were obtained by numerically solving for
$y$ and numerically computing the derivative of $f$ and $s$. The
solid black lines indicate the $|\mathcal{J}|^{-1}$ scaling, which agrees very
well with the numerical results.

\section{Conclusions}

The increasing interest in developing a comprehensive thermodynamic framework for studying complex system, including the process of self-organization, is driven by several recent developments:
theoretical advances in stochastic thermodynamics \cite{Seifert2012} which enable rigorous quantitative analysis of small and mesoscale systems; technological advances that enable measurement of thermodynamic
quantities of such systems~\cite{Berut2015,Nisoli2018,SpinIce2018}; and a fusion of information-theoretic, computation-theoretic and statistical-mechanical approaches for analyzing energy-efficiency of information processing devices \cite{Wolpert2019}.

We have introduced a measure of thermodynamic efficiency of interactions in self-organizing systems, which quantifies the change in the order in the system per unit of the work done/extracted due to the changes in control parameters.
We have shown that this quantity peaks at the critical regime, by explicitly deriving it for the exactly solvable Curie-Weiss model --- a paradigmatic model of second-order phase transitions. Quasi-static perturbation in both control parameters, the interaction strength between spins and the externally applied magnetic field, have been considered, and both protocols have been shown to lead to divergence of the efficiency of interactions at criticality. 

Our work aims to support systematic thermodynamic studies of self-organization in complex systems. In the future, it would be interesting to examine models exhibiting other types of phase transitions in a broad class of dynamical systems, including econo- and socio-computation~\cite{Bouchaud2013,Cancho2003,Cancho2007,Prokopenko2010,Salge2015,Liar2019}, and determine whether they also maximally efficient at criticality. It will also be important to extend the analysis to the protocols that drive the system out-of-equilibrium. We believe that an approach to self-organization incorporating the thermodynamic efficiency will also help 
in clarifying the fundamental relationship between the structure of a complex system
and its collective behavior and function~\cite{Newman2003}, as well as support efforts to systematically
control and guide the dynamics of complex systems \cite{RevModPhys.88.035006,Daniels2017}.

\bibliographystyle{unsrt}
\bibliography{refs}

\end{document}